\pdfoutput=1

\PassOptionsToPackage{x11names,svgnames}{xcolor}

\documentclass{SciPost}
\hypersetup{
    colorlinks,
    linkcolor={red!50!black},
    citecolor={blue!50!black},
    urlcolor={blue!80!black}
}

\usepackage{comment}

\usepackage[bitstream-charter]{mathdesign}
\DeclareSymbolFont{usualmathcal}{OMS}{cmsy}{m}{n}
\DeclareSymbolFontAlphabet{\mathcal}{usualmathcal}

\fancypagestyle{SPstyle}{
\fancyhf{}
\lhead{\colorbox{scipostblue}{\bf \color{white} ~SciPost Physics Codebases }}
\rhead{{\bf \color{scipostdeepblue} ~Submission }}

\fancyfoot[C]{\textbf{\thepage}}
}

\usepackage{doi}

\usepackage{xspace}
\usepackage{graphicx}
\usepackage{afterpage}
\usepackage{lineno}
\usepackage{tikz}
\usetikzlibrary{arrows,shapes,backgrounds,fit,positioning,calc}
\definecolor{irn-bru}{RGB}{239,129,0}

\usepackage{subfig}
\usepackage{microtype}
\makeatletter
\g@addto@macro\bfseries{\boldmath}
\makeatother

\usepackage{xpatch}
\usepackage{xparse}

\usepackage{listings}
\lstMakeShortInline|

\usepackage{todonotes}
\makeatletter
\xpretocmd{\todo}{\@bsphack}{}{}
\xapptocmd{\todo}{\@esphack}{}{}
\makeatother

\usepackage{siunitx}
\DeclareSIUnit{\electronvolt}{\text{e\kern-0.15ex V}}
\DeclareSIUnit{\eV}{\electronvolt}
\DeclareSIUnit{\MeV}{\mega\eV}
\DeclareSIUnit{\GeV}{\giga\eV}
\DeclareSIUnit{\TeV}{\tera\kern-0.1ex\eV}
\DeclareSIUnit{\ifb}{\femto\barn\tothe{-1}}

\NewDocumentCommand{\rivet}{o}{\text{R\protect\scalebox{0.8}{IVET}\IfNoValueTF{#1}{}{\,#1}}\xspace}
\NewDocumentCommand{\contur}{o}{\text{C\protect\scalebox{0.8}{ONTUR}\IfNoValueTF{#1}{}{\,#1}}\xspace}
\NewDocumentCommand{\yoda}{o}{\text{Y\protect\scalebox{0.8}{ODA}\IfNoValueTF{#1}{}{\,#1}}\xspace}
\newcommand{\spey}{S\protect\scalebox{0.8}{PEY}\xspace}

\makeatletter
\let\@oldto\to
\renewcommand{\to}{\ensuremath{\@oldto}\xspace}
\renewcommand{\to}{\ensuremath{\@oldto}\xspace}

\makeatother

\renewcommand{\vec}[1]{\mathbf{#1}}

\NewDocumentCommand{\varcoord}{o}{\ensuremath{\IfNoValueTF{#1}{\vec{\omega}}{\omega^{(#1)}}}\xspace}

\NewDocumentCommand{\bincoord}{o}{\ensuremath{\IfNoValueTF{#1}{\vec{\theta}}{\theta^{(#1)}}}\xspace}

\NewDocumentCommand{\unbincoord}{o}{\ensuremath{\IfNoValueTF{#1}{y}{y}}\xspace} 

\usepackage{placeins}

\hypersetup{hidelinks,
  pdfauthor={Andy Buckley, Jonathan Butterworth, Joseph Egan, Christian Gutschow, Sihyun Jeon, Martin Habedank, Tomasz Procter, Peng Wang, Yoran Yeh, Luzhan Yue}
  pdftitle={Constraints On New Theories Using Rivet : Contur version 3 release note}
}

\begin{document}
\pagestyle{SPstyle}

\begin{center}
  \color{scipostdeepblue}
  \Large
  \textbf{Constraints On New Theories Using Rivet : \contur version 3 release note}
\end{center}

\begin{center}
  A.~G.~Buckley$^1$,
  J.~M.~Butterworth$^{2,*}$,
  J.~C.~Egan$^2$,
  C.~G\"utschow$^{2,3}$,
  M.~Habedank$^1$,
  S.~Jeon$^4$,
  T.~Procter$^5$,
  P.~Wang$^2$,
  Y.~Yeh$^2$,
  L.~Yue$^2$
\end{center}

\begin{center}
  \itshape
  $^1$ School of Physics \& Astronomy, University of Glasgow,\\ University~Place, G12~8QQ, Glasgow, UK\\
  $^2$ Department of Physics \& Astronomy, University College London,\\ Gower~Street, WC1E~6BT, London, UK\\
  $^3$ Centre for Advanced Research Computing, University College London,\\ Gower~Street, London, WC1E~6BT, UK\\
  $^4$ Boston University, 590 Commonwealth Avenue, Boston, Massachusetts 02215, USA \\
  $^5$ Jagiellonian University, \L{}ojasiewicza 11, 30-348 Krakow; Poland.
\end{center}

\begin{center}
\today
\end{center}


\section*{Abstract}
The \contur toolkit exploits \rivet and its library of more than a thousand energy-frontier differential
cross-section measurements from the Large Hadron Collider to allow rapid limit-setting and consistency checks for
new physics models. In this note we summarise the main changes in the new \contur[3] major release
series. These include additional statistical treatments, efficiency improvements, new plotting utilities
and many new measurements and Standard Model predictions.

\clearpage

\begin{centering}
  \small
  \textbf{Other contributors since \contur[2]}\\
  A.~Ajitsaria,
A.~Awnux,
D.~Baig,
S.~Bray,
E.~F.~Butterworth,
M.~Cullingworth,
Z.~Huang,
H.~Hussain,
Y.~Li,
M.~Liu,
M.~Mangat,
S.~Mao,
P.~Mucha,
K.~O'Reilly,
R.~Reine,
S.~Rest,
X.~Wang

\end{centering}

\section{Introduction}
\label{sec:intro}

The \contur toolkit~\cite{Butterworth:2016sqg,Buckley:2021neu} exploits
the analysis library of the 
\rivet toolkit~\cite{Bierlich:2024vqo,Bierlich:2019rhm,Buckley:2010ar}
to set constraints on new beyond the Standard Model (BSM) physics models.
It uses the (large) subset of analyses in \rivet which correspond to
energy-frontier particle-level measurements, usually of differential cross sections
in final-state kinematic variables. These measurements are primarily made
by the ATLAS and CMS collaborations, with a small number from LHCb, and
they are all preserved in HEPData~\cite{Maguire:2017ypu}, from where \rivet accesses them.
The current version of \contur makes use of measurements from 150 different publications,
including a total of more than 1000 measurements.

The \contur workflow involves simulating the full final state predicted
by the new physics model under consideration, and applying the particle-level selection
for the measurements as encoded in \rivet.
The input to \contur is thus the histograms produced by \rivet in the \yoda~\cite{Buckley:2023xqh} format.
Histograms of the events surviving these selections are injected
into the measurement on top of the Standard Model (SM) prediction. Injecting negative contributions
is also possible for generators which can evaluate interference terms; it is also
possible, though of course more time-consuming, to resimulate the entire SM+BSM contribution and compare
that to the data. \contur evaluates whether the BSM contribution improves or degrades the
agreement with data. It reports by default an exclusion at 95\% and 68\% confidence level, as well
as optionally a preferred signal strength, $\hat{\mu}$, and minimum and maximum allowed signal strengths
at 95\% confidence.

\contur[3] is a new major release series, intended and required for use with \rivet[4]. It adds significant new functionality
as well as an enriched library of measurements and SM predictions.

\section{Statistics}
\label{sec:stats}

SM predictions have now been added for all the most impactful measurements.
The primary \contur statistical evaluation now uses these. The previous main analysis method, which
involved injecting potential signals directly on top of the data as described in Ref.~\cite{Buckley:2021neu},
is now turned off by default.

Covariance matrices for the experimental uncertainties are now included where available, allowing a proper
treatment of correlations.

As discussed in Ref.~\cite{GAMBIT:2023yih}, \contur can now use streamed input, and output lower-level
statistical information, so that it can interoperate efficiently with GAMBIT~\cite{GAMBIT:2017yxo}.

As well as the actual and expected exclusions, an estimate can now be made of the eventual sensitivity of High-Luminosity LHC data, first
used and described in Ref.~\cite{Butterworth:2024eyr}. This is
a naive estimate obtained by scaling the uncertainties by the square root of the ratio of the integrated luminosity of each measurement to
the nominal 3~ab$^{-1}$ target, and is likely to underestimate the eventual reach, as discussed for example in Ref.~\cite{Belvedere:2024wzg},
but can nevertheless be a useful indicator.

By default, \contur just evaluates whether or not a given BSM prediction is excluded by the data, using a $\chi^2$ based
likelihood ratio~\cite{Buckley:2021neu}. However, \contur[3] can now optionally make use of the \spey~\cite{Araz:2023bwx} package to perform
a fit of the signal strength $\mu$. The best fit signal strength, ($\hat{\mu}$), and the upper and lower bounds
on the signal strength $\mu$ at 95\% confidence
are output\footnote{First used and described in Refs~\cite{ATLAS:2025oiy,Butterworth:2025asm}.}.
The exclusions given in this case are based upon the ratio to the likelihood at $\hat{\mu}$.
Thus \contur can now be used to determine the maximum allowed cross section for a given
BSM signature. It also means that should a given BSM model improve the description of the data, \contur can give an indication of
evidence \textit{for} that model, rather than just lack of evidence against it.

\section{Plotting improvements}
\label{sec:plotting}

\contur makes extensive use of the Python interfaces to both \rivet and \yoda.
This has been updated for compatibility with the major release series of the \yoda[2]~\cite{Buckley:2023xqh}
statistical data-analysis library. This includes switching to matplotlib for histogram plotting,
via the \yoda toolkit for writing out Python scripts.

\contur results are written to a SQLite~\cite{sqlite2020hipp} database, with the previous pickle format now deprecated and
turned off by default.
Plotting of the exclusion heatmaps and contours is now quicker, due to efficiency improvements
in how this database is read. If grids have been generated in more than two parameter dimensions, it is now easy to select slices 
in the other parameters while generating a two-dimensional heatmap.
A legend is also added to the exclusion plots now by default.

\section{New measurements}
\label{sec:smandmeas}

The measurements added since the \contur[2] publication~\cite{Buckley:2021neu} come from 36 publications, listed in Refs.~\cite{CMS:2012pgw,ATLAS:2016anw,ATLAS:2022fnp,CMS:2022ubq,ATLAS:2023dkz,CMS:2023rcv,ATLAS:2016bxw,CMS:2021maw,CMS:2019eih,ATLAS:2018nci,CMS:2020mxy,ATLAS:2022wnf,ATLAS:2023gsl,ATLAS:2021mbt,ATLAS:2016zba,ATLAS:2020ccu,ATLAS:2024hmk,ATLAS:2024tnr,ATLAS:2022nrp,CMS:2021wfx,ATLAS:2022wmu,ATLAS:2020bbn,ATLAS:2024vqf,ATLAS:2022jat,ATLAS:2019rqw,CMS:2021hnp,ATLAS:2019cbr,CMS:2021vhb,ATLAS:2019zci,ATLAS:2023ibp,CMS:2021lxi,ATLAS:2022xfj,ATLAS:2022mlu,ATLAS:2016zkp,CMS:2022woe,ATLAS:2019iaa,ATLAS:2017nei}. They include many which exploit the full Run 2 dataset of the LHC, including new low-cross section process as
well as precise higher statistics measurement for example of top production, missing energy and dileptons.

\section{Conclusions and outlook}
\label{sec:concl}

The new release series of the \contur toolkit operates with the \rivet[4] and \yoda[2] major releases,
and brings significant enhancements in usability,
statistical treatments and data reach.
Although not part of \contur itself, we note the availability of a pre-processing tool which allows \contur to give information
on models involving long-lived particles~\cite{Corpe:2024ntq}, something not previously possible.
The intention is to make this functionality more readily available in future releases of \rivet and \contur.

As with all previous versions, \contur[3] is available for installation from source and as Docker images, as detailed at
\url{https://hepcedar.gitlab.io/contur-webpage/}. Installation and usage of the new and previously existing features is documented
there, as well as via the command-line of \contur scripts where appropriate.

\section{Acknowledgements}
The authors thank the Marie Sklodowska-Curie Innovative Training
Network MCnetITN3 (grant agreement no. 722104) for funding and
providing the scope for discussion and collaboration toward this
work. AGB, JMB and CG acknowledge funding via the STFC experimental
Consolidated Grants programme (grant numbers %
ST/S000887/1 \& 
ST/W000520/1 
and ST/S000666/1) \& 
ST/W00058X/1), 
and the SWIFT-HEP project (grant numbers
ST/V002562/1 
and ST/V002627/1). 
J.C.E., P.W. and Y.Y. are supported by sources including STFC doctoral training landscape grants, the Spreadbury Fund,
the STFC UCL Centre for Doctoral Training in Data Intensive Science (grant ST/W00674X/1), departmental and industry contributions.
AB, JMB, MH and TP acknowledge funding from OpenMAPP project,
via UKRI/EPSRC EP/Y036360/1 and 
National Science Centre, Poland under CHIST-ERA programme (NCN
2022/04/Y/ST2/00186).
Many thanks to J.~Araz, L.~Corpe, A.~Masouminia, O.~Mattelaer, P.~ Richardson, G.~Watt and others,
especially members of the LPCC Reinterpretation forum~\cite{LHCReinterpretationForum:2020xtr} led by S.~Kraml,
for support and useful discussions, and to D.~Bernard for contributions to the example.

\bibliography{contur.bib,contur-anas.bib}

\appendix

\section{Example: Heavy Dark Mesons and Composite Dark Matter}

A composite dark matter model for Gaugephobic and Gaugephilic $SU(N_D) \,\,;\,\, N_D \in \{2,4\}$ was analysed using CONTUR and data as background in \textit{New sensitivity of LHC measurements to Composite Dark Matter Models} \cite{Butterworth:2021jto}.

The following heatmaps show new constraints using the \textbf{Standard Model as background} (SMBG) (black contours) in the SU(2) case. Additionally, the expected exclusion (black dotted contours) and the expected $2 \sigma$ exclusion at the HL-LHC (red dotted line) are shown. The solid and dashed lines represent the 95\% and 68\% exclusion respectively. The HL-LHC exclusions is estimated as described in \cite{Butterworth:2024eyr}.

All plots were made using the Contur 3.0.0 and used available data from $\sqrt{s} =$ 7, 8 and 13 TeV.

\FloatBarrier
\subsection*{Gaugephilic L}

\begin{figure}[h!]
	\centering
	\includegraphics[width=0.9\textwidth]{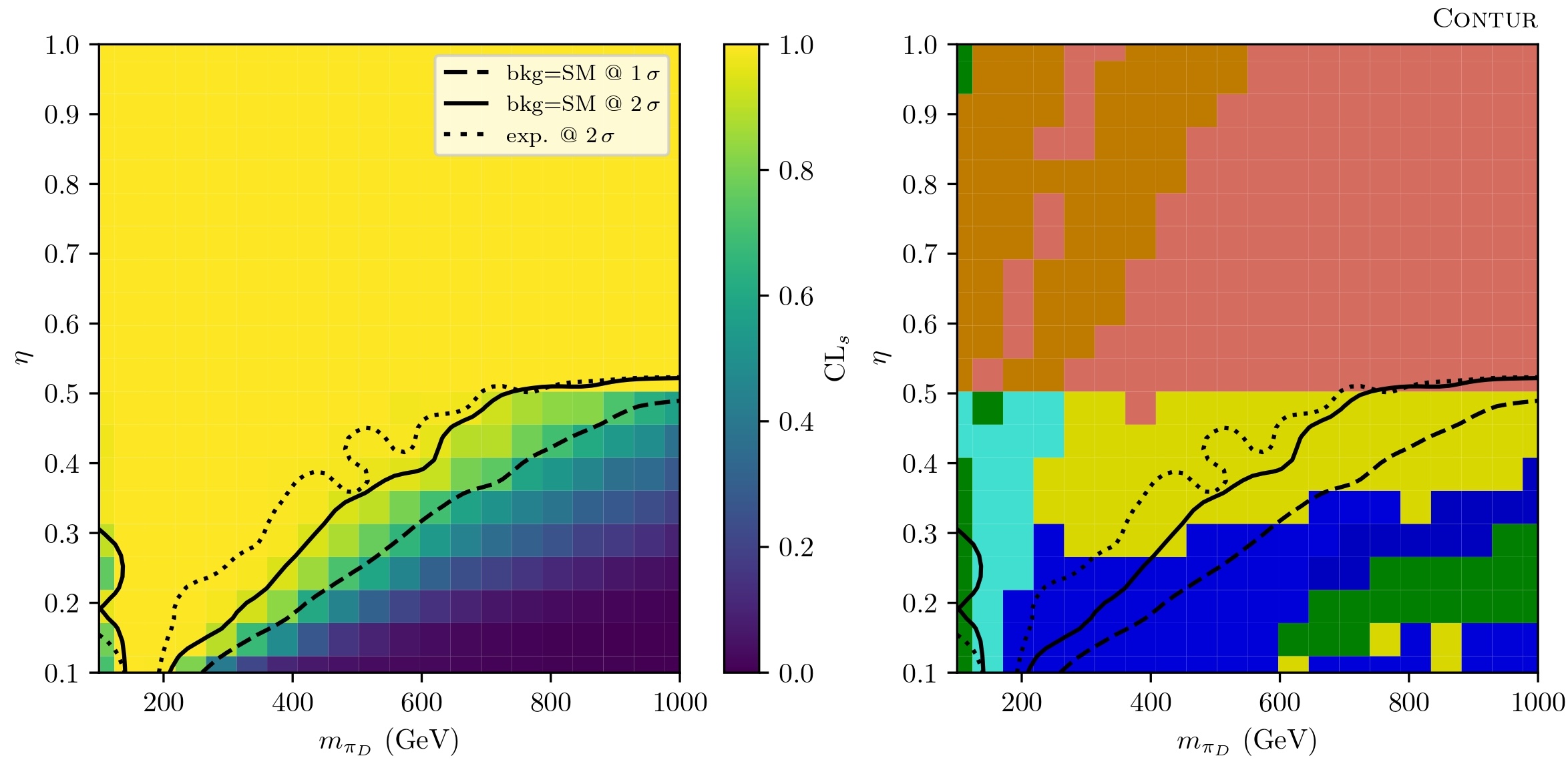} \\
	\includegraphics[width=0.5\textwidth]{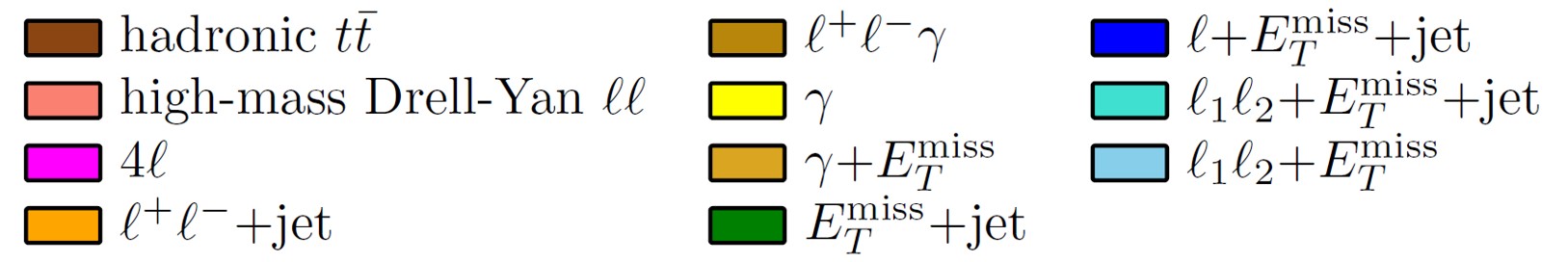}
	\caption{Heatmaps showing exclusion and new dominant pools for the Gaugephilic $SU(2)_L$ model.}
	\label{fig:gaugephilic_su2l}
\end{figure}

\begin{figure}[h!]
	\centering
	\includegraphics[width=0.49\textwidth]{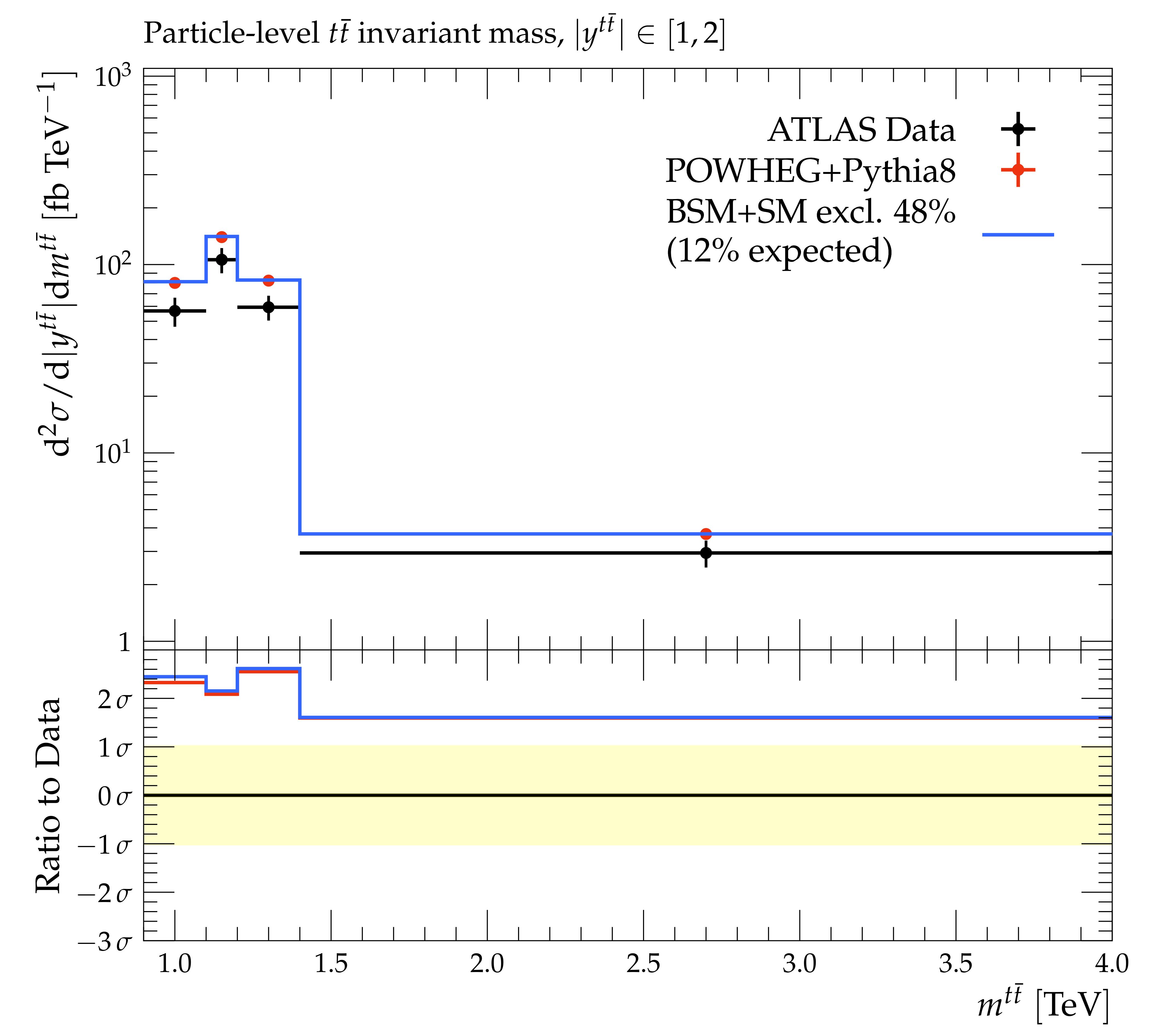}
	\caption{Example histogram for point ($m_{\pi_D}=574$, $\eta=0.36$) in the Gaugephilic $SU(2)_L$ model.}
	\label{fig:histogram1}
\end{figure}

For $\eta > 0.50$, the high-mass Drell-Yan process dominates, as $\rho_D$ is not kinematically allowed to decay into $\pi_D \pi_D$, and therefore decays resonantly to fermions.

From the above figures there is a large disparity between the expected exclusion and the actual exclusion.
This can be explained by the fact that the standard model prediction is already above the data in several measurements,
most notable in the top-antitop measurements and in the ttbb measurement, as shown below. The BSM contribution thus moves the prediction even further from the data.
Contributions from other histograms lead to a 99.57\% exclusion, which is larger than the expected 82.55\% for that reason.

\FloatBarrier
\subsection*{Gaugephobic L}

\begin{figure}[htbp]
	\centering
	\includegraphics[width=0.9\textwidth]{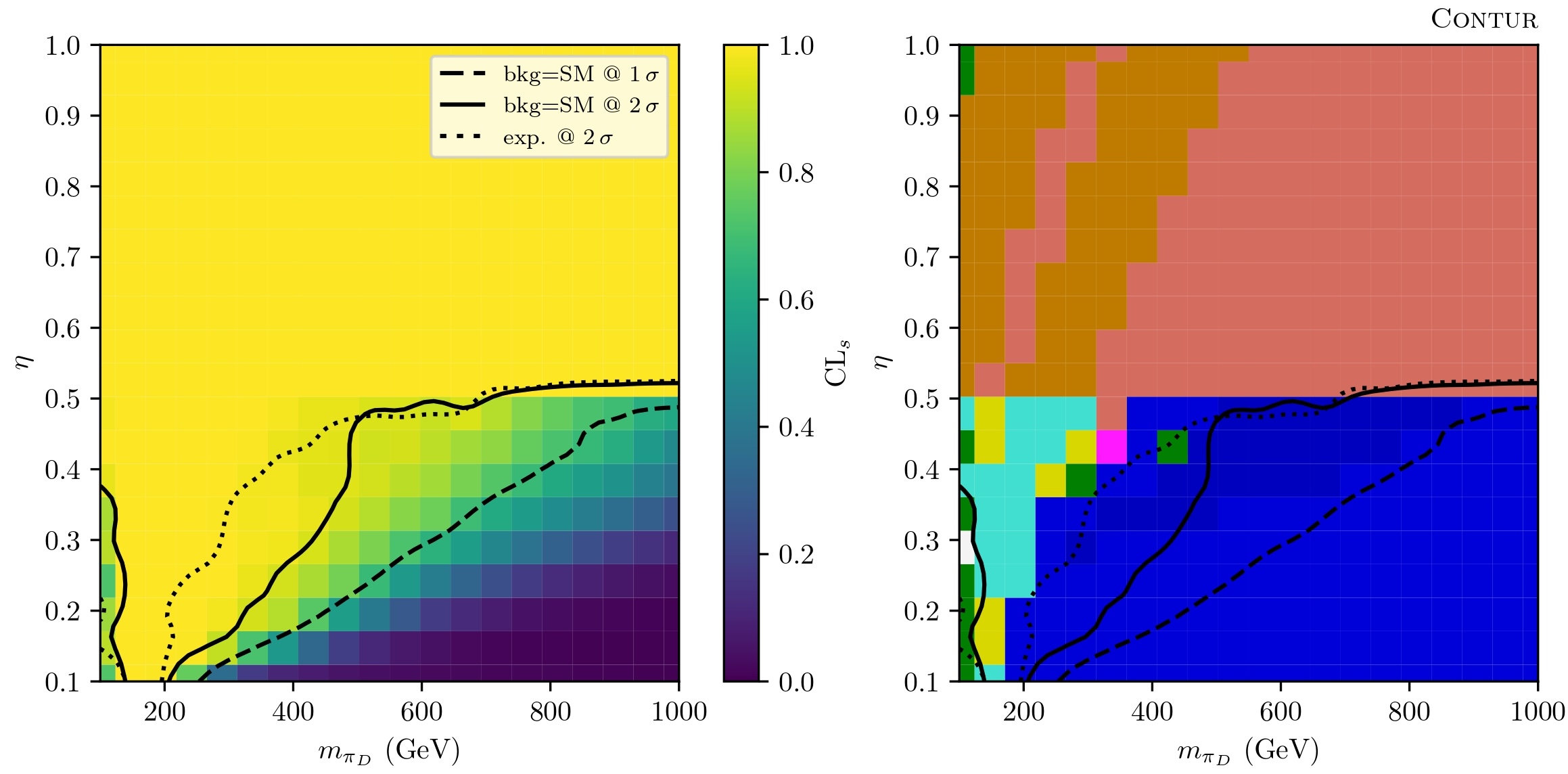}
	\includegraphics[width=0.5\textwidth]{example-plots/legend.jpg}
	\caption{Heatmaps showing new dominant pools for the Gaugephobic $SU(2)_L$ model. The plot on the right has been made as a direct comparison to the one obtained at \cite{ATLAS:2024xbu} (Fig. 13b). As can be seen, the updated limits from Contur are quite similar to the ATLAS search.}
	\label{fig:gaugephobic_su2l}
\end{figure}

Again there is a large disparity between the expected exclusion and the actual exclusion due to the same reason as before.

\FloatBarrier
\subsection*{Gaugephobic R}

\begin{figure}[thb]
	\centering
	\includegraphics[width=0.9\textwidth]{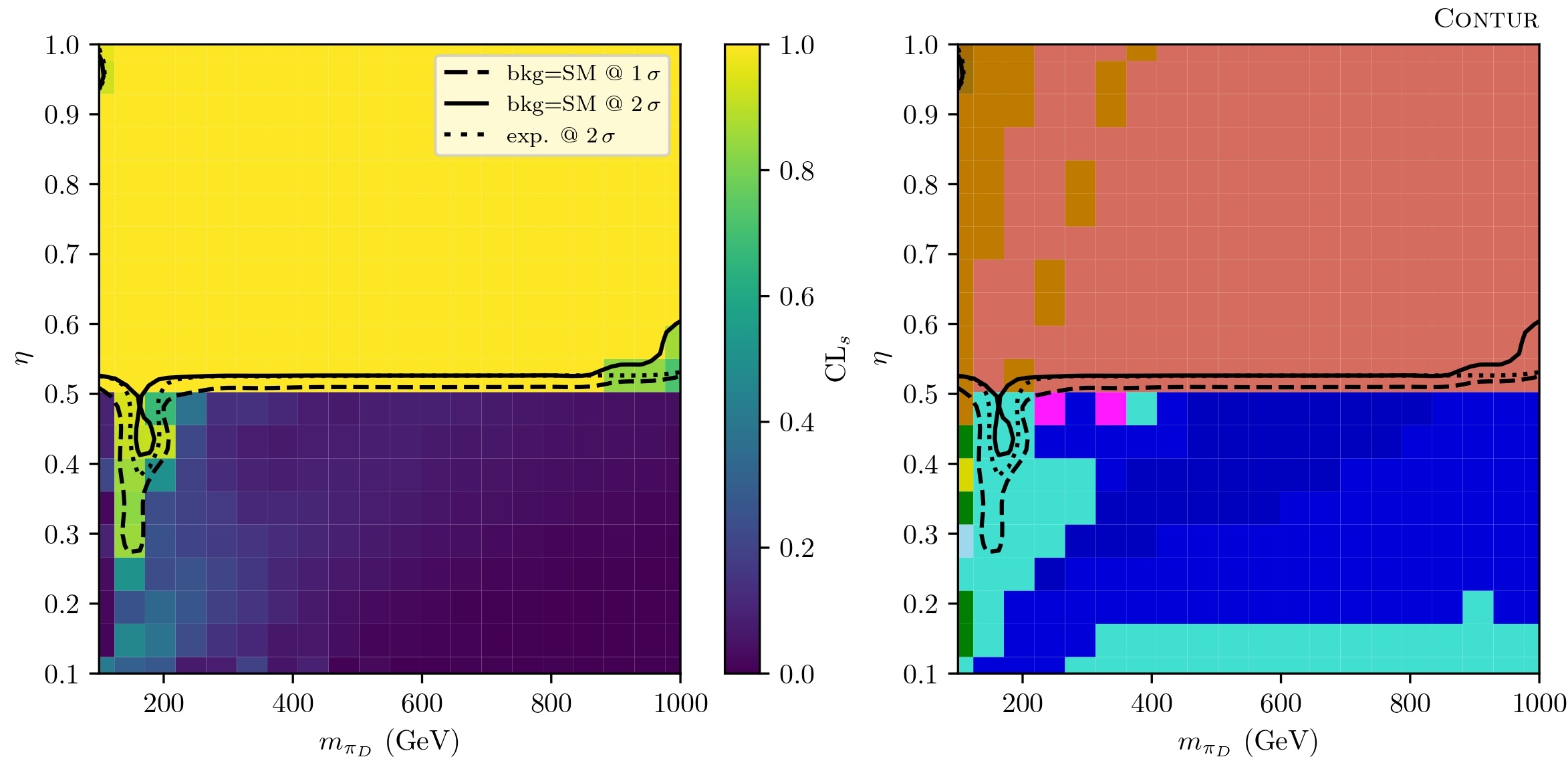}
	\includegraphics[width=0.5\textwidth]{example-plots/legend.jpg}
	\caption{Heatmaps showing new dominant pools for the Gaugephobic $SU(2)_R$ model.}
	\label{fig:gaugephobic_su2r}
\end{figure}

At lower $m_{\pi_D}$, there is a rapid change in CLs between the $\eta < 0.50$ (where $\rho_D$ can decay into $\pi_D \pi_D$), and $\eta > 0.50$ (where it cannot).

\vspace{1.5em}

This example is also available on the website linked above, and model files are also available in the Contur installation in the models directory (under Heavy Dark Mesons).

\end{document}